\documentclass[aps,prx,twocolumn]{revtex4-2}
\usepackage{amsmath}
\usepackage{amssymb}
\usepackage{xcolor}
\usepackage{graphicx}
\usepackage{bm}
\usepackage{bbm}
\usepackage{dcolumn}
\usepackage{graphics}
\usepackage{array}
\usepackage{float}
\usepackage{tikz}
\usepackage{hyperref}
\usepackage{physics}

\begin{document}
\title{Localization pattern of a mobile impurity in the disordered Kitaev chain}
\author{A.V.~Sadovnikov$^{1-3}$ and A.N.~Rubtsov$^{1,4}$}
\affiliation{$^{1}$ Russian Quantum Center, Skolkovo IC, Moscow 121205, Russia} 
\affiliation{$^2$ Moscow Institute of Physics and Technology, 117303 Moscow, Russia}
\affiliation{$^3$ Sechenov University, 119435 Moscow, Russia}
\affiliation{$^4$ Department of Physics, Lomonosov Moscow State University, 119991 Moscow, Russia}

\begin{abstract}
We study a mobile impurity coupled to a Kitaev chain with chemical-potential disorder and ask whether the impurity behavior distinguishes different regimes of the host system. Exact diagonalization calculations for small periodic chains shows that in the deep topological regime the impurity localizes only partially, with a smooth increase of $\mathrm{IPR}_d$, whereas in the deep trivial regime it undergoes a much sharp transition to nearly single-site localization. For open chains at strong interaction, DMRG shows edge-localized impurity density near the Kitaev sweet spot. With increasing chemical potential, the impurity weight spreads into the bulk and eventually becomes almost uniform. We explain the edge preference analytically from the Majorana-dimer structure: a bulk impurity rearranges two neighboring dimers, while an edge impurity affects only one. Disorder competes with this clean edge bias and can pin the impurity in the bulk. Thus, the impurity is sensitive to the regime of the host system, although we do not find a strict one-to-one correspondence between the impurity localization pattern and the host topology. Instead, the disorder-averaged behavior suggests only an indirect correlation between impurity localization and the underlying phase of the chain.

\end{abstract}

\maketitle

\section{Introduction}

One-dimensional topological superconductors provide a minimal setting where a gapped bulk can coexist with Majorana boundary degrees of freedom. The paradigmatic example is the Kitaev chain, in which spinless fermions with effective $p$-wave pairing support zero-energy Majorana modes at the ends of an open chain \cite{kitaev2001unpaired}. Although idealized, this model is closely related to several experimental platforms, including proximitized semiconductor nanowires \cite{lutchyn2010majorana,oreg2010helical,mourik2012signatures,deng2012anomalous}, magnetic-atom chains on superconductors \cite{nadjperge2014observation}, and artificial few-site Kitaev chains in quantum-dot devices \cite{dvir2023realization,bordin2025enhanced}.

Mobile impurities provide another way to probe quantum many-body backgrounds. In topological systems, impurity observables can be sensitive to the structure of the host medium. This idea has been explored in the context of topological polarons \cite{grusdt2016interferometric,grusdt2019topological}, impurities in Chern-insulating backgrounds \cite{camachoguardian2019dropping,pimenov2021topological}, mobile impurities in one-dimensional topological lattice models \cite{valiente2019flat}, and polarons in topological superfluids \cite{qin2019polaron}. These studies show that impurities can carry information about topological environments, although the relation between impurity response and topology is generally model- and observable-dependent.

Disorder is also essential in realistic topological systems. It can modify phase boundaries, localize bulk states, generate low-energy in-gap features, and complicate the interpretation of experimental signatures. In some cases, disorder can even induce nontrivial topology, as in topological Anderson insulators \cite{li2009topological,meier2018observation}. In Majorana platforms, disorder effects have been studied in Kitaev chains and semiconductor nanowires \cite{pan2021disorder,dassarma2021disorder}, including interacting disordered Kitaev chains where moderate disorder or interactions can stabilize topological order while strong disorder suppresses it \cite{gergs2016topological}.

In this work we study a single mobile impurity locally coupled to a disordered Kitaev chain. Rather than reconstructing the disordered topological phase diagram of the host, we focus on how chemical-potential disorder is converted into an effective real-space background for the impurity. We analyze the impurity IPR and density profile using exact diagonalization and DMRG, and interpret the numerical results using analytically solvable limits. The manuscript is organized as follows. In Sec.~\ref{sec:model}, we introduce the model and observables. In Sec.~\ref{sec:results}, we present the numerical results for periodic and open boundary conditions. In Sec.~\ref{sec:solvable_limits}, we discuss limiting cases that clarify the effective-landscape mechanism. Finally, in Sec.~\ref{sec:discussion}, we summarize the physical picture emerging from the numerical and analytical results.

\section{Model}\label{sec:model}

In our previous paper, we studied a Kitaev chain locally coupled to a heavy mobile impurity~\cite{sadovnikov2026mobile}. Here, we extend this model by adding on-site disorder to the chemical potential of the Kitaev chain. The total Hamiltonian is
\begin{equation}\label{syst}
    H = H_{\text{Kit}} + H_\text{imp} + H_\text{int},
\end{equation}
where
\begin{equation}\label{kit}
    H_\text{Kit} =
    \sum_i
    \left(
    -t_c c^\dagger_i c_{i+1}
    + \Delta c^\dagger_i c_{i+1}^\dagger
    + h.c.
    - \mu_i c^\dagger_i c_i
    \right)
\end{equation}
is the Hamiltonian of the Kitaev chain,
\begin{equation}\label{imp}
    H_\text{imp} =
    -t_d\sum_i
    \left(
    d^\dagger_i d_{i+1}
    +
    h.c.
    \right)
\end{equation}
describes the mobile impurity, and
\begin{equation}\label{int}
    H_\text{int}
    =
    U\sum_i
    c^\dagger_i c_i d^\dagger_i d_i
\end{equation}
is the local density-density interaction between the impurity and the Kitaev-chain fermions.

Here, $c_i^\dagger$ creates a fermion in the Kitaev chain, while $d_i^\dagger$ creates the impurity fermion. The parameters $t_c$ and $t_d$ are the hopping amplitudes of the Kitaev-chain fermions and the impurity, respectively, $\Delta$ is the superconducting pairing amplitude, and $U$ is the local impurity--fermion interaction.

Throughout this paper, all Hamiltonian parameters are measured in units of the Kitaev-chain hopping amplitude $t_c$. We set
\begin{equation*}
    t_c=1, \qquad \Delta=1, \qquad t_d=0.1,
\end{equation*}
unless stated otherwise. Thus, the pairing amplitude and the impurity hopping are kept fixed, while the average chemical potential $\mu$, the disorder strength $W$, and the interaction strength $U$ are varied.

We consider both periodic and open boundary conditions. In the periodic case, the sums in Eqs.~\eqref{kit} and~\eqref{imp} are understood with $c_{N+1}=c_1$ and $d_{N+1}=d_1$. In the open-boundary case, the hopping and pairing sums terminate at $N-1$.

In the clean limit, $\mu_i=\mu$, the Kitaev chain has two phases. For $|\mu|<2t_c$, the system is in the topological phase and supports Majorana zero modes localized near the edges under open boundary conditions. For $|\mu|>2t_c$, the system is in the trivial phase and has no such edge zero modes. Since we use $t_c=1$, this criterion becomes
\begin{equation*}
    \begin{cases}
        |\mu|<2, & \text{topological phase}, \\
        |\mu|>2, & \text{trivial phase}.
    \end{cases}
\end{equation*}

In the disordered case, the on-site chemical potential is taken in the form
\begin{equation*}
    \mu_i = \mu + w_i,
\end{equation*}
where $\mu$ is the average chemical potential and $w_i$ are independent random variables drawn from the uniform distribution
\begin{equation*}
    w_i \sim \mathrm{Unif}[-W,W].
\end{equation*}
The parameter $W$ therefore controls the strength of the diagonal disorder in the Kitaev chain.

The impurity is described by a simple tight-binding Hamiltonian with hopping amplitude $t_d$. For periodic boundary conditions, its single-particle dispersion is
\begin{equation*}
    \varepsilon_k^d = -2t_d\cos k .
\end{equation*}
With our choice $t_d=0.1$, this gives $\varepsilon_k^d=-0.2\cos k$, so the impurity has a narrow bandwidth equal to $4t_d=0.4$.

Due to the superconducting pairing term in $H_\text{Kit}$, the number of $c$-fermions is not conserved. By contrast, the impurity number is conserved. Throughout this work, we restrict ourselves to the single-impurity sector,
\begin{equation*}
    \sum_{j=1}^{N}\langle d_j^\dagger d_j\rangle = 1.
\end{equation*}

To characterize the real-space localization of the impurity, we use the impurity inverse participation ratio,
\begin{equation}\label{eq:IPR}
    \text{IPR}_d
    =
    \sum_{i=1}^{N}
    \langle n_i^d\rangle^2,
    \qquad
    n_i^d=d_i^\dagger d_i .
\end{equation}
Since the impurity number is fixed to one, this quantity satisfies
\begin{equation*}
    \frac{1}{N}
    \leq
    \text{IPR}_d
    \leq
    1 .
\end{equation*}
The lower bound corresponds to a spatially uniform impurity density,
$\langle n_i^d\rangle=1/N$, while the upper bound corresponds to the impurity density being concentrated on a single site.

We emphasize that Eq.~\eqref{eq:IPR} is a real-space impurity-density IPR, not a many-body inverse participation ratio in Fock space. Thus, in finite systems, a large value of $\text{IPR}_d$ should be interpreted as strong spatial concentration of the impurity density rather than as direct evidence for an Anderson-localized phase in the thermodynamic sense.


\section{Results}\label{sec:results}

\begin{figure*}[t]
\centering
\includegraphics[width=0.49\linewidth]{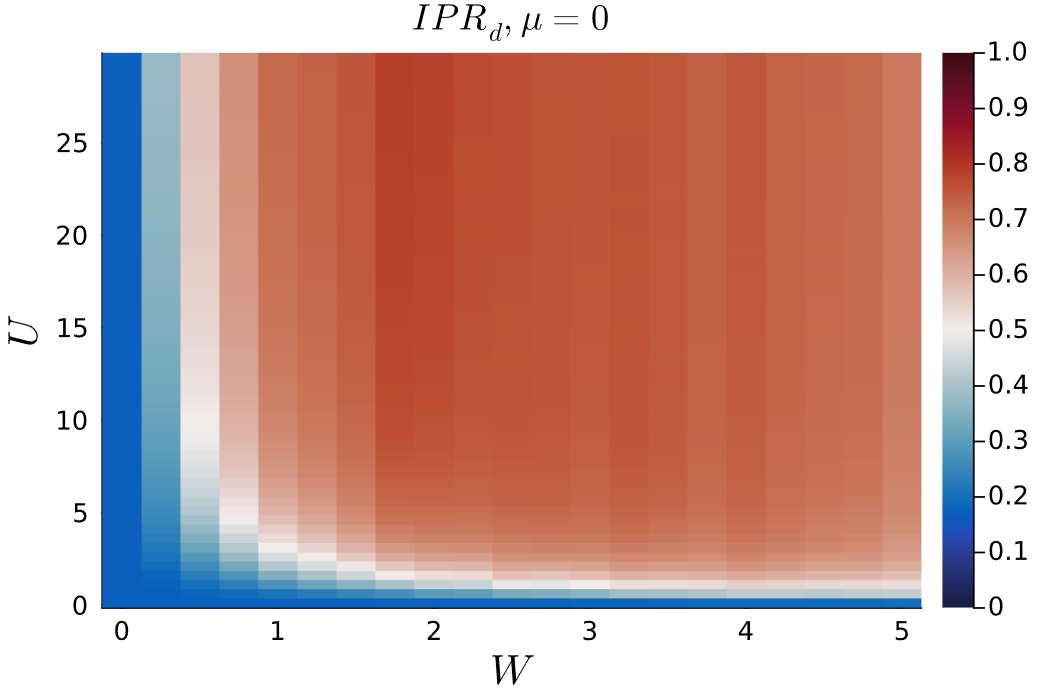}
\includegraphics[width=0.49\linewidth]{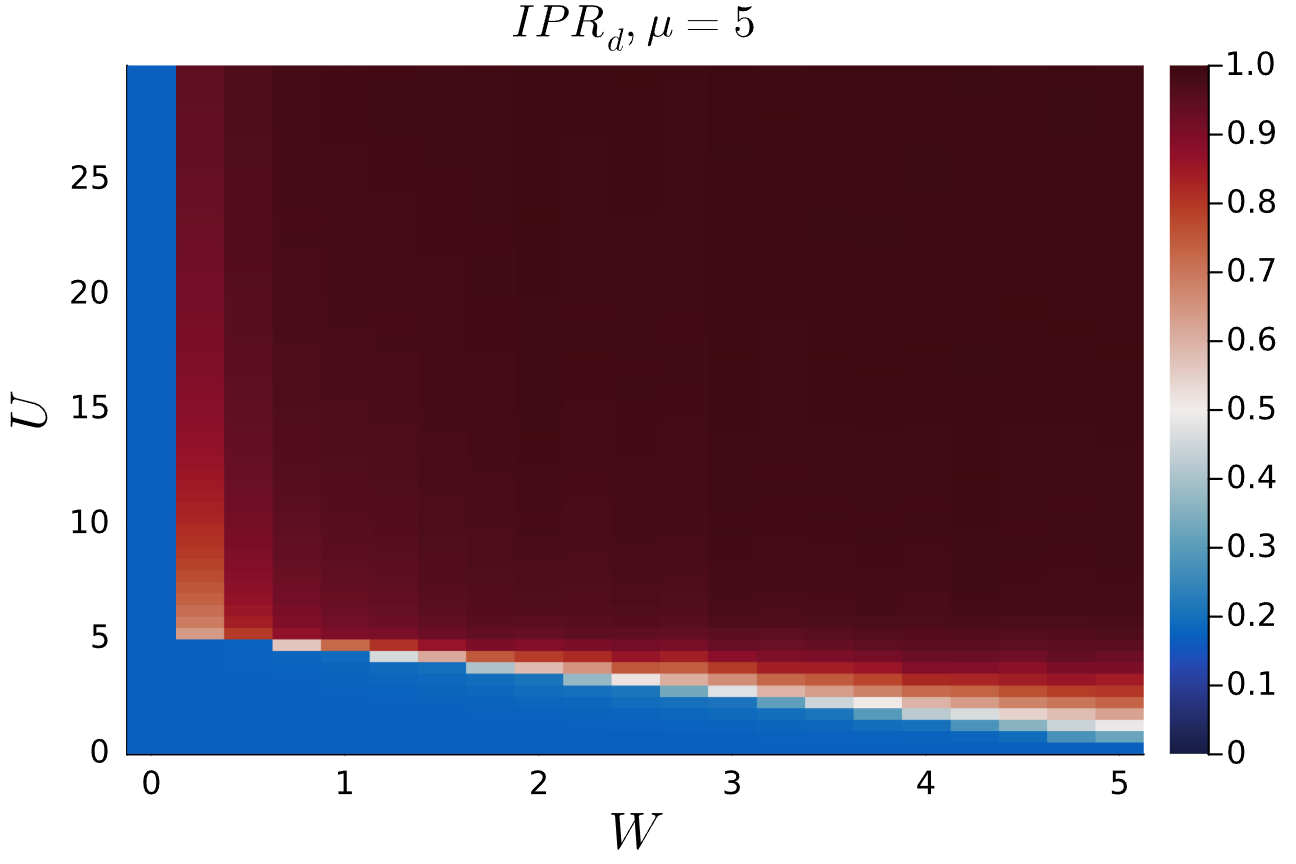}
\caption{
Disorder-averaged real-space impurity ${\text{IPR}}_d$, Eq.~\eqref{eq:IPR}, for periodic boundary conditions with $N=6$ sites. The left panel corresponds to the topological bulk regime of the clean Kitaev chain, $\mu=0$, while the right panel corresponds to the trivial regime, $\mu=5$.
}
\label{fig:IPR_PBC}
\end{figure*}

\textit{Periodical boundaries:}  Figure~\ref{fig:IPR_PBC} shows the disorder-averaged impurity-density IPR, Eq.~\eqref{eq:IPR}, obtained by exact diagonalization for periodic boundary conditions for $N=6$ sites. The two panels correspond to $\mu=0$ and $\mu=5$, respectively.

It is important to note that periodic boundary conditions remove physical edge, so the data will characterize how the bulk properties of the disordered superconductor affect the impurity and the must not be connected to the Majorana edge modes.

For $\mu=0$, the impurity IPR increases smoothly with both the interaction strength $U$ and the disorder amplitude $W$. In the region with interaction  $U \leq 5$ and weak disorder $W\leq 0.75$, $\text{IPR}_d$ remains close to its lower finite-size value, indicating that the impurity density is broadly distributed over the chain. For $W>0.75$ the IPR almost immediately reaches it maximum value with increasing $U$. The maximum IPR value is about $\sim 0.6-0.7$ that indicates no single-site localization in the deep topological regime for a small system.

For $\mu=5$, the behavior is qualitatively different. There is a parameter region, shown in blue color, where the impurity is fully delocalized. Increasing $W$ lowers the critical interaction strength $U_c$, which separates the delocalized impurity state from the fully localized state shown in dark red color. The transition is sharp around $U\sim 5$ for $W\leq 1.5$, but becomes smoother at larger $W$. At $W=5$, the sharp transition almost disappears and start to look like a smooth crossover from the weakly localized state (light blue color) to the fully localized state.

\textit{Open boundaries:} At weak interaction we do not observe qualitative changes compared with the clean-chain results of Ref.~\cite{sadovnikov2026mobile}: after disorder averaging, the impurity-density profile remains essentially unchanged. The effect of disorder becomes more visible in the strong-interaction regime.

\begin{figure*}[t]
    \centering
    \includegraphics[width=0.49\linewidth]{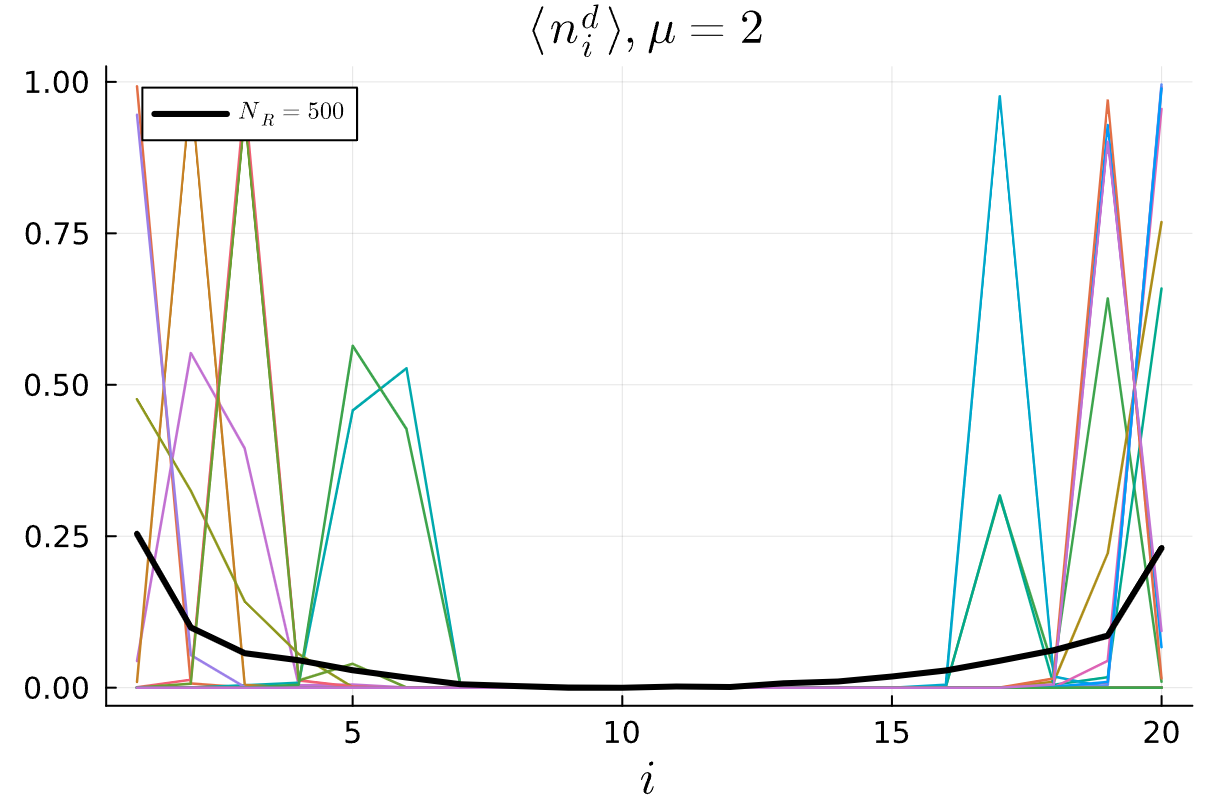}
    \includegraphics[width=0.49\linewidth]{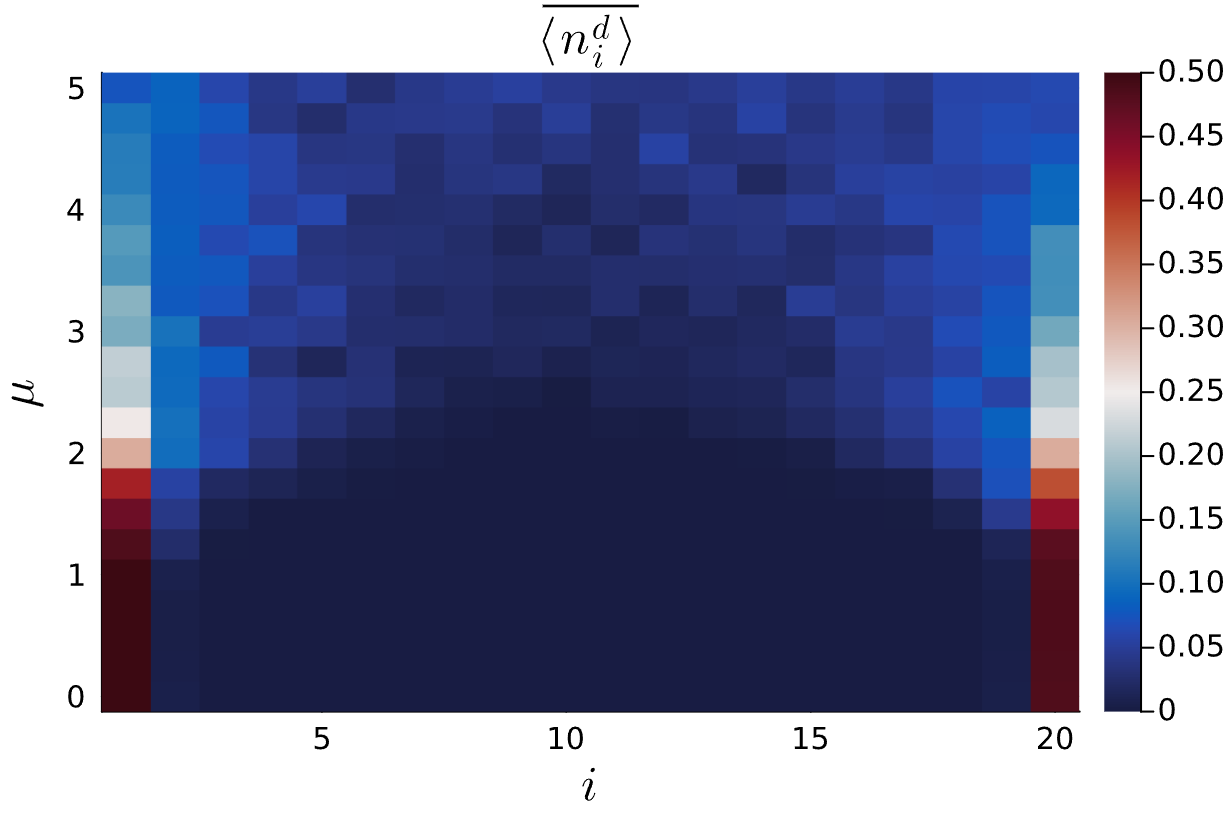}
    \caption{Distribution of the impurity along the open chain at high interaction strength. Left: illustration of the impurity density $n^d_i$ for $N_R=20$ of disorder realizations without disorder averaging at $\mu=2$; black solid line represents the disorder-avereged density at $\mu=2$. Right: disorder-averaged impurity density $\overline{\langle n_i^d\rangle}$ with $N_R=500$. The calculation was performed using DMRG at $U=10$ and $W=0.25$.
}
    \label{fig:nd_obc}
\end{figure*}

Figure~\ref{fig:nd_obc} shows the impurity density for open boundary conditions at strong interaction, $U=10$, and disorder strength $W=0.25$. The left panel illustrates the impurity density profiles at fixed chemical potential $\mu=2$ for $N_R=20$ individual disorder realizations. The colored curves show that the impurity position is strongly sample dependent: in many realizations the largest weight remains close to one of the boundaries, while in some realizations the impurity is pinned at an interior site. The black solid line shows the disorder-averaged density at the same value of $\mu$. The averaged profile still shows larger weight near the edges, but the sharp peaks present in individual realizations are smoothed out by averaging.

The right panel of Fig.~\ref{fig:nd_obc} shows the disorder-averaged impurity density $\overline{\langle n_i^d\rangle}$ as a function of the site index $i$ and the average chemical potential $\mu$, averaged over $N_R=500$ disorder realizations. At chemical potential $\mu\leq 1.25$, the impurity density is strongly concentrated near the two boundaries, while the bulk density is strongly suppressed. This behavior is consistent with the edge preference derived analytically at the Kitaev sweet spot in Sec.~\ref{subsec: dissordered sweet spot}: placing the impurity at the boundary produces a smaller reconstruction of the Majorana-dimer pattern than placing it in the bulk. At larger $\mu$ the impurity density starts to spread from the edges into the bulk, and at $\mu=5$ looks almost uniformly distributed along the chain. The bulk weight grows smoothly rather than through a sharp jump, reflecting the averaging over realization-dependent impurity positions. 

It is important to note one point: although the impurity localization may appear to be related to the Majorana zero modes of the host system, we did not find a direct connection between them. They can be indirectly related by the fact that in the deep topological phase it is more probably to find the impurity near the on of the edges, while in the deep trivial phase it can be found everywhere including the edges.

\section{Exactly solvable limits}\label{sec:solvable_limits}

We consider open boundary conditions throughout this section. In order to develop an analytic understanding of the impurity localization problem, it is useful to isolate parameter regimes in which the electronic sector can be diagonalized exactly after fixing the impurity position. This provides a controlled effective description of the impurity motion and clarifies which mechanisms originate from the clean open-chain boundary structure and which are induced by disorder.

Strong coupling regime $U \gg 1$ allows to study the system at fixed impurity position $j$. The electronic Hamiltonian becomes
\begin{equation}\label{hamiltonian_electronic}
H_c^{(j)} = H_c + U n_j^c.
\end{equation}
Its ground-state energy,
\begin{equation*}
\varepsilon_j = E_0(H_c^{(j)}),
\end{equation*}
defines the effective potential felt by the impurity. In the regime of small $t_d$, the low-energy impurity dynamics is controlled by the effective tight-binding Hamiltonian
\begin{equation*}
H_{\text{eff}}
=
\sum_{j=1}^{N}\varepsilon_j \dyad{j}{j}
-
t_d\sum_{j=1}^{N-1}
\left(
\dyad{j}{j+1}+\dyad{j+1}{j}
\right),
\end{equation*}
up to corrections due to the nontrivial overlap of many-body electronic ground states for different impurity positions. The exactly solvable limits discussed below provide explicit study of $\varepsilon_j$'s contribution in the localization.

\subsection{Clean topological sweet spot for the immobile impurity}

At the topological sweet spot,
\begin{equation*}
t_c=\Delta,
\qquad
\mu_j=0,
\end{equation*}
it is convenient to introduce Majorana operators
\begin{equation*}
a_j= c_j^\dagger + c_j,
\qquad
b_j=i(c^\dagger_j-c_j)
\qquad
i a_j b_j = 2n_j^c-1.
\end{equation*}
Then the fixed-impurity electronic Hamiltonian Eq.\eqref{hamiltonian_electronic} becomes
\begin{equation}\label{hamiltonian_majorana}
H_c^{(j)}
=\frac{i}{2}
\left[
U a_j b_j + 2t_c\sum_{\ell=1}^{N-1} b_\ell a_{\ell+1}
\right]
= \frac{i}{2}\sum_{\ell,m=1}^{N} a_\ell M_{\ell m}^{(j)} b_m,
\end{equation}
with $M^{(j)}=2t_c\,S+U E_{jj}$, $S_{\ell m}=\delta_{\ell,m+1}$ and $(E_{jj})_{\ell m}=\delta_{\ell j}\delta_{mj}$. If $s_\alpha^{(j)}$ are the singular values of $M^{(j)}$, then the fixed-impurity ground-state energy is
\begin{equation*}
\varepsilon_j=\frac{U}{2}-\frac12\sum_{\alpha=1}^{N}s_\alpha^{(j)}.
\end{equation*}

For $j=1$ (and analogously for $j=L$), the matrix $M^{(1)}$ is lower bidiagonal with one nontrivial boundary column, so that its singular values are $\sqrt{U^2+4t_c^2}$, together with $L-2$ values equal to $2t_c$ and one zero mode. Hence
\begin{equation}\label{eq:epsilon_edge}
{
\varepsilon_{\text{edge}}
=
\frac{U}{2}
-(L-2)t_c
-\frac12\sqrt{U^2+4t_c^2}.
}
\end{equation}
For $2\le j\le N-1$, the only nontrivial local block is
\begin{equation*}
B=
\begin{pmatrix}
2t_c & U\\
0 & 2t_c
\end{pmatrix},
\end{equation*}
whose singular values $s_\pm$ satisfy
\begin{equation*}
s_+s_-=4t_c^2,
\text{ }
s_+^2+s_-^2=U^2+8t_c^2, \text{ }
s_++s_-=\sqrt{U^2+16t_c^2}.
\end{equation*}
Since the remaining $L-3$ nonzero singular values are equal to $2t_c$, one finds
\begin{equation}\label{ed:epsilon_bulk}
\varepsilon_{\text{bulk}} = \frac{U}{2} -(L-3)t_c -\frac{1}{2}\sqrt{U^2+16 t_c^2}.
\end{equation}
Therefore the difference of the bulk and edge energies $\Delta^{(0)} = \varepsilon_{\text{bulk}}-\varepsilon_{\text{edge}}$ equals to
\begin{equation}\label{eq: Delta^0}
\Delta^{(0)}
=
t_c+\frac12\sqrt{U^2+4t_c^2}-\frac12\sqrt{U^2+16t_c^2}>0,
\end{equation}
because this inequality is equivalent to $\sqrt{U^2+4t_c^2}>2t_c$, which holds for all $U>0$. Thus
\begin{equation*}
{
\varepsilon_{\text{edge}}<\varepsilon_{\text{bulk}}
\qquad \text{for all } U>0,
}
\end{equation*}
which provides an exact microscopic derivation within the sweet-spot limit.

From a physical point of view, this result means that a fully localized impurity effectively divides the chain into two parts. In the Majorana representation, the $U$-term competes with two neighboring Majorana dimers, $(b_{j-1},a_j)$ and $(b_j,a_{j+1})$, and locally rearranges the dimerization pattern. The energetically favorable configuration is therefore the edge-localized one: at the edge, the impurity affects only one dimer and produces a smaller condensation-energy cost than in the bulk.

\subsection{Disordered sweet spot}\label{subsec: dissordered sweet spot}

We now add weak chemical-potential disorder to the sweet-spot Hamiltonian,
\begin{equation*}
    \mu_i=w_i,
    \qquad
    |w_i|\ll U,
    \qquad
    \overline{w_i}=0 .
\end{equation*}
For a fixed impurity position $j$, the disorder enters the Kitaev Hamiltonian Eq.\eqref{kit} as $ V_{\rm dis} = -\sum_i w_i n_i^c$. We denote the corresponding ground-state energy by $\varepsilon_j(\mathbf w)$. Making use of the Feynman-Hellmann theorem,
\begin{equation*}
    \frac{\partial \varepsilon_j}{\partial w_i} = - \langle n_i^c\rangle_j ,
\end{equation*}
and therefore, to first order perturbation theory in disorder,
\begin{equation}\label{eq:FH_disorder}
    \varepsilon_j(\mathbf w) = \varepsilon_j(0) - \sum_i w_i \rho_i^{(j)}
    + O(w^2),
    \quad
    \rho_i^{(j)} = \langle n_i^c\rangle_j^{(0)} ,
\end{equation}
where $\varepsilon_j(0) =\varepsilon_\text{edge}$ Eq.\eqref{eq:epsilon_edge} for $j=1,N$ and $\varepsilon_j(0) =\varepsilon_\text{bulk}$ Eq.\eqref{ed:epsilon_bulk} for $2\le j\le N-1$,.

The density profile $\rho_i^{(j)}$ can be obtained directly from the singular-value decomposition used in the clean Hamiltonian Eq.\eqref{hamiltonian_majorana}. Let
\begin{equation*}
    Q^{(j)}
    =
    M^{(j)}
    \left[
    (M^{(j)})^T M^{(j)}
    \right]^{-1/2}_{+}
\end{equation*}
be the partial polar factor, where the inverse is understood on the subspace of nonzero singular values. The Majorana correlator gives
\begin{equation*}
    \langle i a_i b_i\rangle_j
    =
    -Q_{ii}^{(j)},
    \qquad
    \rho_i^{(j)}
    =
    \frac12-\frac12 Q_{ii}^{(j)} .
    \label{eq:density_Q}
\end{equation*}
Thus Eq.\eqref{eq:FH_disorder} becomes
\begin{equation*}
    \varepsilon_j(\mathbf w)
    =
    \varepsilon_j(0)
    -
    \frac12\sum_i w_i
    +
    \frac12\sum_i w_i Q_{ii}^{(j)}
    +
    O(w^2).
    \label{eq:energy_Q}
\end{equation*}
The first disorder term is independent of the impurity position and therefore drops out from all energy differences.

At the sweet spot the diagonal part of the polar factor is especially simple. For an edge impurity $j=1 \text{ or }N$
\begin{equation*}
    Q_{ii}^{(\text{1})}
    =
    \delta_{i\text{1}}
    \frac{U}{\sqrt{U^2+4t_c^2}} = \delta_{i\text{1}}A_\text{e},
    \label{eq:Q_edge}
\end{equation*}
whereas for a bulk impurity $2\le j\le N-1$,
\begin{equation*}
    Q_{ii}^{(j)}
    =
    \delta_{ij}
    \frac{U}{\sqrt{U^2+16t_c^2}} = \delta_{ij} A_\text{b}.
    \label{eq:Q_bulk}
\end{equation*}
Therefore, one can obtain the energy difference $\Delta_j(\mathbf{w}) = \varepsilon_j(\mathbf{w}) - \varepsilon_\text{1}(\mathbf{w})$:
\begin{equation*}
    \Delta_j(\mathbf w)
    =
    \Delta^{(0)}
    +
    \frac12
    \left(
        A_{\rm b} w_j
        -
        A_{\rm e} w_\text{1}
    \right)
    +
    O(w^2),
\end{equation*}

where $\Delta^{(0)}$ is given by the Eq.\eqref{eq: Delta^0}. If the difference $\Delta_j(\mathbf{w}) \geq 0$ the impurity is energetically favorable to localize at the edge of the system. 

The disorder term is most unfavorable to the edge when $w_j$ is negative and $w_\text{1}$ is positive. Therefore, for all disorder realizations satisfying $|w_i|\le W$,
\begin{equation*}
    \frac12
    \left(
        A_{\rm b} w_j
        -
        A_{\rm e} w_\text{1}
    \right)
    \ge
    -
    \frac{W}{2}
    \left(
        A_{\rm b}+A_{\rm e}
    \right).
\end{equation*}
Hence
\begin{equation*}
    \Delta_j(\mathbf w)
    \ge
    \Delta^{(0)}
    -
    \frac{W}{2}
    \left(
        A_{\rm b}+A_{\rm e}
    \right)
    +
    O(w^2).
    \label{eq:disordered_gap_bound}
\end{equation*}
To leading order in weak disorder, the edge configuration is therefore stable for every realization if
\begin{equation*}
    W
    <
    W_c^{(1)},
    \qquad
    W_c^{(1)}
    =
    \frac{2\Delta^{(0)}}{A_{\rm b}+A_{\rm e}} .
    \label{eq:Wc_sweet_exact}
\end{equation*}
Substituting the explicit expressions, this gives
\begin{equation*}
    W_c^{(1)}
    =
    \frac{
    2t_c+\sqrt{U^2+4t_c^2}-\sqrt{U^2+16t_c^2}
    }{
    \dfrac{U}{\sqrt{U^2+16t_c^2}}
    +
    \dfrac{U}{\sqrt{U^2+4t_c^2}}
    } ,
    \label{eq:Wc_sweet_explicit}
\end{equation*}
and in the strong-coupling limit $U\gg t_c$,
\begin{equation*}
    W_c^{(1)}
    =
    t_c-\frac{3t_c^2}{U}
    +
    O(U^{-2}).
\end{equation*}
Equivalently, the most unfavorable disorder configuration gives the energy difference
\begin{equation}\label{eq:Delta(w)}
    \Delta(\mathbf w)
    \ge t_c - \frac{3t_c^2}{U} -W +O(w^2,U^{-2}).
\end{equation}
Thus, for $|w_i|\le W$, the clean edge preference is perturbatively protected for $W < t_c-\frac{3t_c^2}{U}$ in the large-$U$ limit. A bulk minimum can appear only once the disorder amplitude reaches the clean edge-bulk gap scale.

We now include the impurity hopping $t_d$. Taking the edge energy as the reference point and subtracting $\varepsilon_{\rm e}(\mathbf w)$ into the Hamiltonian Eq.\eqref{hamiltonian_electronic}, we obtain
\begin{equation*}
H_{\rm eff}'=\sum_{j=2}^{N}\Delta_j(\mathbf w)\dyad{j}{j}-t_d\sum_{j=1}^{N-1}\left(\dyad{j}{j+1}+\dyad{j+1}{j}\right),
\end{equation*}
where $\Delta_j(\mathbf w)=\varepsilon_j(\mathbf w)-\varepsilon_{\rm e}(\mathbf w)$.

The most unfavorable bounded disorder configuration for the edge is $w_{\rm e}=W$ and $w_\text{b}=-W$ for the competing bulk site. Using Eq.~\eqref{eq:Delta(w)}, this gives
\begin{equation*}
\Delta_j(\mathbf w)\ge \Delta_{\min}=t_c-\frac{3t_c^2}{U}-W+O(W^2,U^{-2}).
\end{equation*}
Thus a sufficient comparison problem is obtained by replacing all bulk onsite energies by the lower bound $\Delta_{\min}$:
\begin{equation*}
H_{\rm comp}=\Delta_{\min}\sum_{j=2}^{N}\dyad{j}{j}-t_d\sum_{j=1}^{N-1}\left(\dyad{j}{j+1}+\dyad{j+1}{j}\right).
\end{equation*}
After subtracting the constant $\Delta_{\min}$, this is equivalent to a tight-binding chain with an attractive edge potential $-\Delta_{\min}$.

For this boundary problem, an edge state has the form $\psi_j\propto r^{j-1}$. The Schrödinger equations give $r=t_d/\Delta_{\min}$, so localization requires $|r|<1$, or equivalently $\Delta_{\min}>t_d$. Hence a sufficient leading-order condition for edge localization of the mobile impurity is
\begin{equation*}
t_c-\frac{3t_c^2}{U}-W>t_d+O(W^2,U^{-2}),
\end{equation*}
or, up to higher-order corrections,
\begin{equation}
W+t_d<t_c-\frac{3t_c^2}{U}.
\end{equation}

\subsection{Atomic limit}

We now turn to the atomic electronic limit,
\begin{equation*}
t_c=\Delta=0,
\end{equation*}
for which the full Hamiltonian Eq.\eqref{syst} is
\begin{equation*}
H=
-\sum_{j=1}^{N}\mu_j n_j^c
-t_d\sum_{j=1}^{N-1}(d_j^\dagger d_{j+1}+d_{j+1}^\dagger d_j)
+U\sum_{j=1}^{N} n_j^d n_j^c.
\label{eq:atomic_full_compact}
\end{equation*}
Since
\begin{equation*}
[H,n_j^c]=0 \qquad \forall j,
\end{equation*}
the electronic occupations are conserved and may be treated as binary variables $n_j\in\{0,1\}$. For an impurity fixed at site $j$ and $t_d=0$, the energy in a given electronic configuration is
\begin{equation*}
E_j(\mathbf n)=-\sum_{i=1}^{N}\mu_i n_i + U n_j.
\end{equation*}
Minimization over $\mathbf n$ is carried out independently at each site: for $i\neq j$ one obtains $\min(-\mu_i n_i)=-(\mu_i)_+$, whereas at the impurity site $\min[(U-\mu_j)n_j]=-(\mu_j-U)_+$. Hence
\begin{equation*}
E_j^{(0)}
=
-\sum_{i\neq j}(\mu_i)_+-(\mu_j-U)_+.
\label{eq:Ej0_atomic_compact}
\end{equation*}
It is convenient to separate the $j$-independent part by writing
\begin{equation*}
E_j^{(0)}
=
-\sum_{i=1}^{N}(\mu_i)_+
+
\Big[(\mu_j)_+-(\mu_j-U)_+\Big].
\end{equation*}
This motivates the definition
\begin{equation*}
{
\eta_j = (\mu_j)_+-(\mu_j-U)_+,
}
\end{equation*}
so that $E_j^{(0)}=\text{const}+\eta_j$. In other words, $\eta_j$ is the exact local energy cost of placing the impurity at site $j$ after optimizing the electronic occupations. Explicitly,
\begin{equation}
\eta_j=
\begin{cases}
0, & \mu_j\le 0,\\
\mu_j, & 0<\mu_j<U,\\
U, & \mu_j\ge U.
\end{cases}
\label{eq:eta}
\end{equation}
In the clean atomic limit, $\mu_j=\mu$, the energy cost $\eta_j$ is spatially uniform. Therefore, the interaction alone does not select a preferred impurity position in this limit. In other words, the clean atomic limit contains no energetic mechanism for real-space impurity pinning.

The term $\eta_j$ play the role of effective potential in the Hamilotinan Eq.\eqref{hamiltonian_electronic} $\varepsilon_j$
\begin{equation*}
{
H_{\text{eff}}
=
\sum_{j=1}^{N}\eta_j \dyad{j}{j}
-
t_d\sum_{j=1}^{N-1}\left(\dyad{j}{j+1}+\dyad{j+1}{j}\right).
}
\label{eq:Heff_atomic_compact}
\end{equation*}
If $\eta_j$ has a unique global minimum at some interior site $j_*$, separated from the rest by
\begin{equation*}
\gamma=\min_{j\neq j_*}(\eta_j-\eta_{j_*})>0,
\end{equation*}
If $t_d\ll\gamma$, standard non-degenerate perturbation theory gives
\begin{equation*}
\lambda_0
=
\eta_{j_*}
-
t_d^2\left[
\frac{1}{\eta_{j_*-1}-\eta_{j_*}}
+
\frac{1}{\eta_{j_*+1}-\eta_{j_*}}
\right]
+O(t_d^3),
\end{equation*}
and the corresponding wave function is exponentially localized around $j_*$, with amplitudes decaying as $(t_d/\gamma)^r$. Conversely, if $\mu_j>U$ on all sites, then $\eta_j=U$ is flat and no localization mechanism is present.

\section{Discussion}\label{sec:discussion}

We study a mobile impurity locally interacting with the disordered Kitaev chain. The results show that the impurity response is sensitive to the difference between the deep topological and deep trivial regimes, but it is not a direct diagnostic of the topological phase transition. For periodic boundary conditions, the impurity shows strong but incomplete localization at $\mu=0$, while at $\mu=5$ it can undergo a sharp transition to an almost single-site localized state. Thus the trivial regime supports stronger disorder-induced pinning, whereas the topological regime keeps the impurity density more broadly distributed.

For open boundary conditions, the main effect at small $\mu$ is the clean edge preference. At the Kitaev sweet spot this preference has a simple energetic origin: a bulk impurity rearranges two neighboring Majorana dimers, whereas an edge impurity affects only one. This gives $\Delta^{(0)}=\varepsilon_{\rm bulk}-\varepsilon_{\rm edge}=t_c+\frac12\sqrt{U^2+4t_c^2}-\frac12\sqrt{U^2+16t_c^2}>0.$ Weak chemical-potential disorder competes with this bias. For bounded disorder $|w_i|\le W$, the most unfavorable configuration gives, in the large-$U$ limit $\Delta(\mathbf w)\ge t_c-\frac{3t_c^2}{U}-W+O(W^2,U^{-2}).$ For a mobile impurity, the hopping $t_d$ provides an additional delocalizing scale. A sufficient perturbative condition for the edge-localized state to survive is $W+t_d<t_c-\frac{3t_c^2}{U}$, up to higher-order corrections. When this condition is not satisfied, the impurity weight can spread from the edges into the bulk, consistent with the smooth evolution of the disorder-averaged density profile.

In the atomic limit with $t_c=\Delta=0$, the clean system has no preferred impurity position, and pinning appears only from spatial variations of the local energy cost $\eta_j$ governed by the Eq.\eqref{eq:eta}. Thus edge localization near the sweet spot and disorder-induced bulk pinning are distinct microscopic mechanisms.

From an experimental point of view, the mobile impurity considered here should be regarded as an idealized local probe. A genuinely mobile fermionic impurity is natural in cold-atom polaron settings~\cite{Schirotzek2009}. In solid-state platforms, a closer analogue would be a controllable local fermionic level, such as an additional quantum dot, a gate-defined state, or a scanning probe coupled locally to the chain. This makes the present results relevant to recent quantum-dot implementations of minimal Kitaev chains~\cite{Dvir2023,Bordin2025,Bordin2026,SeoaneSouto2023}. In particular, one part of our analysis is performed for a small system size, comparable in spirit to current few-site quantum-dot chains. Even at this level, the impurity response clearly distinguishes the deep topological and deep trivial regimes, showing that local probes can be sensitive to the underlying superconducting background before one reaches the thermodynamic limit.

However, this sensitivity should not be interpreted as a standalone Majorana diagnostic. If the probe response is strongly concentrated near the ends of an open chain in the expected topological regime, this may indicate where Majorana signatures should be searched for. Nevertheless, edge localization of the impurity is not by itself evidence for Majorana zero modes. In our analysis, the edge preference already appears from the clean Majorana-dimer reconstruction energy, and disorder can create competing bulk pinning centers. Since disorder and trivial Andreev bound states can also produce local low-energy signatures similar to Majorana signals~\cite{DasSarma2021,Pan2019}, the impurity profile should be used only as a complementary marker of boundary-dominated behavior, together with independent spectroscopic or nonlocal probes.

\section*{Acknowledgments}
A. Sadovnikov acknowledges support of the Foundation for the Advancement of Theoretical Physics and Mathematics “BASIS”.  

The authors are grateful to Anton Markov and Murod Bahovadinov for the useful comments and discussions.

\bibliography{bibl}

@article{kitaev2001unpaired,
  title={Unpaired Majorana fermions in quantum wires},
  author={Kitaev, A Yu},
  journal={Physics-uspekhi},
  volume={44},
  number={10S},
  pages={131},
  year={2001},
  publisher={IOP Publishing}
}

@article{sadovnikov2026mobile,
  title={Mobile impurity in the Kitaev chain: Phase diagram and signatures of topology},
  author={Sadovnikov, AV and Bahovadinov, MS and Rubtsov, AN and Markov, AA},
  journal={Physical Review B},
  volume={113},
  number={15},
  pages={155146},
  year={2026},
  publisher={APS}
}

@article{valiente2019flat,
  title={Flat band of topological states bound to a mobile impurity},
  author={Valiente, Manuel},
  journal={arXiv preprint arXiv:1907.08215},
  year={2019}
}

@article{lutchyn2010majorana,
  title={Majorana fermions and a topological phase transition in semiconductor-superconductor heterostructures},
  author={Lutchyn, Roman M. and Sau, Jay D. and Das Sarma, S.},
  journal={Physical Review Letters},
  volume={105},
  number={7},
  pages={077001},
  year={2010},
  publisher={American Physical Society},
  doi={10.1103/PhysRevLett.105.077001}
}

@article{oreg2010helical,
  title={Helical liquids and Majorana bound states in quantum wires},
  author={Oreg, Yuval and Refael, Gil and von Oppen, Felix},
  journal={Physical Review Letters},
  volume={105},
  number={17},
  pages={177002},
  year={2010},
  publisher={American Physical Society},
  doi={10.1103/PhysRevLett.105.177002}
}

@article{mourik2012signatures,
  title={Signatures of Majorana fermions in hybrid superconductor-semiconductor nanowire devices},
  author={Mourik, V. and Zuo, K. and Frolov, S. M. and Plissard, S. R. and Bakkers, E. P. A. M. and Kouwenhoven, L. P.},
  journal={Science},
  volume={336},
  number={6084},
  pages={1003--1007},
  year={2012},
  publisher={American Association for the Advancement of Science},
  doi={10.1126/science.1222360}
}

@article{deng2012anomalous,
  title={Anomalous zero-bias conductance peak in a Nb--InSb nanowire--Nb hybrid device},
  author={Deng, M. T. and Yu, C. L. and Huang, G. Y. and Larsson, M. and Caroff, P. and Xu, H. Q.},
  journal={Nano Letters},
  volume={12},
  number={12},
  pages={6414--6419},
  year={2012},
  publisher={American Chemical Society},
  doi={10.1021/nl303758w}
}

@article{nadjperge2014observation,
  title={Observation of Majorana fermions in ferromagnetic atomic chains on a superconductor},
  author={Nadj-Perge, Stevan and Drozdov, Ilya K. and Li, Jian and Chen, Hua and Jeon, Sangjun and Seo, Jungpil and MacDonald, Allan H. and Bernevig, B. Andrei and Yazdani, Ali},
  journal={Science},
  volume={346},
  number={6209},
  pages={602--607},
  year={2014},
  publisher={American Association for the Advancement of Science},
  doi={10.1126/science.1259327}
}

@article{dvir2023realization,
  title={Realization of a minimal Kitaev chain in coupled quantum dots},
  author={Dvir, Tom and Wang, Guanzhong and van Loo, Nick and Liu, Chun-Xiao and Mazur, Grzegorz P. and Bordin, Alberto and ten Haaf, Sebastiaan L. D. and Wang, Ji-Yin and van Driel, David and Zatelli, Francesco and Li, Xiang and Malinowski, Filip K. and Kouwenhoven, Leo P.},
  journal={Nature},
  volume={614},
  number={7948},
  pages={445--450},
  year={2023},
  publisher={Nature Publishing Group},
  doi={10.1038/s41586-022-05585-1}
}

@article{bordin2025enhanced,
  title={Enhanced Majorana stability in a three-site Kitaev chain},
  author={Bordin, Alberto and others},
  journal={Nature Nanotechnology},
  volume={20},
  pages={XXXX--XXXX},
  year={2025},
  publisher={Nature Publishing Group},
  doi={10.1038/s41565-025-01894-4}
}

@article{grusdt2016interferometric,
  title={Interferometric measurements of many-body topological invariants using mobile impurities},
  author={Grusdt, Fabian and Abanin, Dmitry and Demler, Eugene},
  journal={Nature Communications},
  volume={7},
  pages={11994},
  year={2016},
  publisher={Nature Publishing Group},
  doi={10.1038/ncomms11994}
}

@article{grusdt2019topological,
  title={Topological polarons, quasiparticle invariants, and their detection in one-dimensional symmetry-protected phases},
  author={Grusdt, Fabian and Yao, Norman Y. and Demler, Eugene A.},
  journal={Physical Review B},
  volume={100},
  number={7},
  pages={075126},
  year={2019},
  publisher={American Physical Society},
  doi={10.1103/PhysRevB.100.075126}
}

@article{camachoguardian2019dropping,
  title={Dropping an impurity into a Chern insulator: A polaron view on topological matter},
  author={Camacho-Guardian, A. and Goldman, N. and Massignan, P. and Bruun, G. M.},
  journal={Physical Review B},
  volume={99},
  number={8},
  pages={081105},
  year={2019},
  publisher={American Physical Society},
  doi={10.1103/PhysRevB.99.081105}
}

@article{pimenov2021topological,
  title={Topological transport of mobile impurities},
  author={Pimenov, D. and Camacho-Guardian, A. and Goldman, N. and Massignan, P. and Bruun, G. M. and Goldstein, M.},
  journal={Physical Review B},
  volume={103},
  number={24},
  pages={245106},
  year={2021},
  publisher={American Physical Society},
  doi={10.1103/PhysRevB.103.245106}
}

@article{qin2019polaron,
  title={Polaron in a $p+ip$ Fermi topological superfluid},
  author={Qin, Fang and Cui, Xiaoling and Yi, Wei},
  journal={Physical Review A},
  volume={99},
  number={3},
  pages={033613},
  year={2019},
  publisher={American Physical Society},
  doi={10.1103/PhysRevA.99.033613}
}

@article{li2009topological,
  title={Topological Anderson Insulator},
  author={Li, Jian and Chu, Rui-Lin and Jain, Jainendra K. and Shen, Shun-Qing},
  journal={Physical Review Letters},
  volume={102},
  number={13},
  pages={136806},
  year={2009},
  publisher={American Physical Society},
  doi={10.1103/PhysRevLett.102.136806}
}

@article{meier2018observation,
  title={Observation of the topological Anderson insulator in disordered atomic wires},
  author={Meier, Eric J. and An, Fangzhao Alex and Gadway, Bryce},
  journal={Science},
  volume={362},
  number={6417},
  pages={929--933},
  year={2018},
  publisher={American Association for the Advancement of Science},
  doi={10.1126/science.aat3406}
}

@article{pan2021disorder,
  title={Disorder effects on Majorana zero modes: Kitaev chain versus semiconductor nanowire},
  author={Pan, Haining and Das Sarma, S.},
  journal={Physical Review B},
  volume={103},
  number={22},
  pages={224505},
  year={2021},
  publisher={American Physical Society},
  doi={10.1103/PhysRevB.103.224505}
}

@article{dassarma2021disorder,
  title={Disorder-induced zero-bias peaks in Majorana nanowires},
  author={Das Sarma, Sankar and Pan, Haining and Sau, Jay D.},
  journal={Physical Review B},
  volume={103},
  number={19},
  pages={195158},
  year={2021},
  publisher={American Physical Society},
  doi={10.1103/PhysRevB.103.195158}
}

@article{gergs2016topological,
  title={Topological order in the Kitaev/Majorana chain in the presence of disorder and interactions},
  author={Gergs, Niklas M. and Fritz, Lars and Schuricht, Dirk},
  journal={Physical Review B},
  volume={93},
  number={7},
  pages={075129},
  year={2016},
  publisher={American Physical Society},
  doi={10.1103/PhysRevB.93.075129}
}

@article{Schirotzek2009,
  title = {Observation of Fermi Polarons in a Tunable Fermi Liquid of Ultracold Atoms},
  author = {Schirotzek, A. and Wu, C.-H. and Sommer, A. and Zwierlein, M. W.},
  journal = {Physical Review Letters},
  volume = {102},
  pages = {230402},
  year = {2009},
  doi = {10.1103/PhysRevLett.102.230402}
}

@article{Dvir2023,
  title = {Realization of a Minimal Kitaev Chain in Coupled Quantum Dots},
  author = {Dvir, T. and others},
  journal = {Nature},
  volume = {614},
  pages = {445--450},
  year = {2023},
  doi = {10.1038/s41586-022-05585-1}
}

@article{Bordin2025,
  title = {Enhanced Majorana Stability in a Three-Site Kitaev Chain},
  author = {Bordin, A. and others},
  journal = {Nature Nanotechnology},
  year = {2025},
  doi = {10.1038/s41565-025-01894-4}
}

@article{Bordin2026,
  title = {Probing Majorana Localization of a Phase-Controlled Three-Site Kitaev Chain with an Additional Quantum Dot},
  author = {Bordin, A. and others},
  journal = {Nature Communications},
  year = {2026},
  doi = {10.1038/s41467-026-68897-0}
}

@article{SeoaneSouto2023,
  title = {Probing Majorana Localization in Minimal Kitaev Chains through a Quantum Dot},
  author = {Seoane Souto, Rub{\'e}n and Tsintzis, Athanasios and Leijnse, Martin and Danon, Jeroen},
  journal = {arXiv:2308.14751},
  year = {2023},
  doi = {10.48550/arXiv.2308.14751}
}

@article{DasSarma2021,
  title = {Disorder-Induced Zero-Bias Peaks in Majorana Nanowires},
  author = {Das Sarma, Sankar and Pan, Haining},
  journal = {Physical Review B},
  volume = {103},
  pages = {195158},
  year = {2021},
  doi = {10.1103/PhysRevB.103.195158}
}

@article{Pan2019,
  title = {Physical Mechanisms for Zero-Bias Conductance Peaks in Majorana Nanowires},
  author = {Pan, Haining and Sau, Jay D. and Das Sarma, Sankar},
  journal = {arXiv:1910.11413},
  year = {2019},
  doi = {10.48550/arXiv.1910.11413}
}

\end{document}